\documentclass[aps,pra,showpacs,groupedaddress]{revtex4}
\usepackage{amssymb,bm}
\usepackage[dvips]{graphicx}

\begin{document}
\bibliographystyle{apsrev}

\title{Polarizing mechanisms for stored $p$ and $\bar p$ beams interacting
 with a polarized target}

\author{A.I. Milstein  }
\email{A.I.Milstein@inp.nsk.su}
\author{V.M.Strakhovenko}
\email{V.M.Strakhovenko@inp.nsk.su} \affiliation{Budker Institute
of Nuclear Physics, 630090 Novosibirsk, Russia}

\date{\today}

\begin{abstract}
Kinetics of the polarization buildup at the interaction of stored
protons (antiprotons) with a polarized target is considered. It is
demonstrated that for small scattering angles, when a projectile
remains in the beam, the polarization buildup is completely due to
the spin-flip transitions. The corresponding cross sections turn
out to be negligibly small for a hydrogen gas target as well as
for  a pure electron target. For the latter, the filtering
mechanism also does not provide  a noticeable beam polarization.

\end{abstract}

\maketitle

\section{Introduction}

 At present, the use of a polarized hydrogen gas target to
polarize stored antiprotons is widely discussed (see, e.g.,
\cite{Rathmann05} and literature therein). The modifications of
the method first proposed in \cite{Csonka68} are considered. The
original idea of spin filtering  is based on the dependence of the
scattering cross section on the mutual orientation of the target
and projectile proton spins. Then the protons having opposite spin
projections on the direction of the target polarization go out
from the beam with the different rates.  As a result, the beam
acquires a polarization.

The method suggested in  \cite{Csonka68} has been realized in the
experiment \cite{Haeberli93}, where  23-MeV stored protons
scattered on an internal gas target of the polarized hydrogen
atoms. In 90 minutes,  the intensity of the beam was 5\% of the
initial one and the polarization degree amounted to 2.4\%. In
\cite{Haeberli93}, the rate of polarization buildup due to
filtering mechanism was also estimated theoretically,  taking into
account only strong $pp$ interaction. This estimation is
noticeably different from the experimental result. The explanation
of this disagreement is proposed in \cite{Meyer94}. In that paper
the importance of the interference of the Coulomb amplitude and
the spin-dependent part of the hadronic amplitude is emphasized.
Under conditions of the experiment \cite{Haeberli93}, this effect
diminishes the corresponding cross section more than by 40 \%,
thereby, improving essentially  agreement with the experiment. Our
calculations confirm the estimation  of this effect obtained in
\cite{Meyer94}.

The remaining difference between the experimental and theoretical
results, which is rather small, is explained in \cite{Meyer94} by
 two new mechanisms. Both mechanisms are related to scattering at the angle
$\vartheta$ smaller than the acceptance angle $\theta_{acc}\ll 1$,
where protons remain in the beam. The first effect, suggested in
\cite{MeyerHorowitz94}, is due to the interaction of a projectile
with the polarized electrons of the hydrogen gas target. The
second one, considered in \cite{Meyer94}, is due to scattering on
polarized protons of the target at  $\vartheta<\theta_{acc}$. The
estimation of these two effects, made in \cite{Meyer94}, gives the
contributions  to the rate of polarization buildup close in the
absolute values but of the opposite sign. The magnitude of each
effect is comparable with that corresponding to filtering effect.
The result accounting for all three contributions agrees very well
with the experiment. Basing on the results of
\cite{MeyerHorowitz94}, a method to polarize an antiproton beam
was suggested in \cite{Rathmann05}. The idea of Ref.
\cite{Rathmann05} is in the use of a hydrogen gas target with the
high electron and low  proton polarizations.

In the present paper, we demonstrate that the consideration of
both new effects performed in \cite{Meyer94,MeyerHorowitz94} is
not correct. We show that, for scattering at
$\vartheta<\theta_{acc}$ (a projectile remains in the beam), the
polarization buildup is completely due to spin-flip transitions.
For $\vartheta\ll 1$, a noticeable contribution to the
spin-dependent part of the differential cross section appears as a
result of interference between the spin-independent Coulomb
amplitude and spin-dependent amplitude. This is because of  the
singularity of the  Coulomb amplitude at small scattering angles
($\propto 1/\vartheta^2$). Evidently, such interference is absent
in  scattering  with the spin flip of a projectile, and the
corresponding cross section is negligibly small. The formulas used
in \cite{Meyer94,MeyerHorowitz94} for the description of their two
effects correspond to such interference and, therefore, are
irrelevant to the kinetics of polarization.

\section{Kinetics of polarization}

We consider a beam of particles with the densities in the momentum
space $f_+({\bm p},t)$ and $f_-({\bm p},t)$  , where subscripts
correspond to the  spin projections $\pm 1/2$ on a quantization
axis. Let $W_{\sigma_f \sigma_i}({\bm p}_f,{\bm p}_i)$ be the
probability of the transition  from the state with the momentum
${\bm p}_i$ and polarization $\sigma_i$ to the state with the
momentum ${\bm p}_f$ and polarization $\sigma_f$. Then we have for
the densities $f_\pm$ the conventional kinetic equation:
\begin{eqnarray}\label{eq:f}
\frac{\partial}{\partial t}f_+({\bm p},t)&=&-\int d{\bm
p}'[W_{++}({\bm p}',{\bm
p})+W_{-+}({\bm p}',{\bm p})]f_+({\bm p},t)\nonumber\\
&&+\int\limits_{{\bm p}'\in \Gamma} d{\bm p}'[W_{++}({\bm p},{\bm
p}')f_+({\bm p}',t)+W_{+-}({\bm p},{\bm p}')f_-({\bm p}',t)]\, ,
\nonumber\\
\frac{\partial}{\partial t}f_-({\bm p},t)&=&-\int d{\bm
p}'[W_{--}({\bm p}',{\bm
p})+W_{+-}({\bm p}',{\bm p})]f_-({\bm p},t)\nonumber\\
&&+\int\limits_{{\bm p}'\in \Gamma} d{\bm p}'[W_{-+}({\bm p},{\bm
p}')f_+({\bm p}',t)+W_{--}({\bm p},{\bm p}')f_-({\bm p}',t)]\, .
\end{eqnarray}
Here ${\bm p}'\in \Gamma$ means that the momentum ${\bm p}'$
belongs to the beam momentum space (the angle between the momentum
${\bm p}'$ and the beam axis is less than $\theta_{acc}$). Taking
in Eq.(\ref{eq:f}) the integral over ${\bm p}\in \Gamma$, we
obtain
\begin{eqnarray}\label{eq:f1}
\frac{\partial}{\partial t}\int\limits_{{\bm p}\in \Gamma} d{\bm
p}[f_+({\bm p},t)-f_-({\bm p},t)]&=&-\int\limits_{{\bm p}'\notin
\Gamma} d{\bm p}'\!\!\int\limits_{{\bm p}\in \Gamma} d{\bm
p}\,[W_{++}({\bm p}',{\bm
p})+W_{-+}({\bm p}',{\bm p})]f_+({\bm p},t)\nonumber\\
&&+\int\limits_{{\bm p}'\notin \Gamma} d{\bm
p}'\!\!\int\limits_{{\bm p}\in \Gamma} d{\bm p}\,[W_{--}({\bm
p}',{\bm p})+W_{+-}({\bm p}',{\bm p})]f_-({\bm p},t)\nonumber\\
&&+2\int\limits_{{\bm p}'\in \Gamma} d{\bm
p}'\!\!\int\limits_{{\bm p}\in \Gamma} d{\bm p}\,[W_{+-}({\bm
p}',{\bm
p})f_-({\bm p},t)-W_{-+}({\bm p}',{\bm p})f_+({\bm p},t)]\, ,\nonumber\\
\frac{\partial}{\partial t}\int\limits_{{\bm p}\in \Gamma} d{\bm
p}[f_+({\bm p},t)+f_-({\bm p},t)]&=&-\int\limits_{{\bm p}'\notin
\Gamma} d{\bm p}'\!\!\int\limits_{{\bm p}\in \Gamma} d{\bm
p}\,[W_{++}({\bm p}',{\bm
p})+W_{-+}({\bm p}',{\bm p})]f_+({\bm p},t)\nonumber\\
&&-\int\limits_{{\bm p}'\notin \Gamma} d{\bm
p}'\!\!\int\limits_{{\bm p}\in \Gamma} d{\bm p}\,[W_{--}({\bm
p}',{\bm p})+W_{+-}({\bm p}',{\bm p})]f_-({\bm p},t)\, .
\end{eqnarray}
As seen from Eq.(\ref{eq:f1}), in the  term where both momenta
${\bm p}$ and ${\bm p}'$ belong to $\Gamma$, only the spin-flip
probabilities are present. In other words, scattering without loss
of particles may lead to a beam polarization solely due  to the
spin-flip transitions. Due to phase space cooling, the
distributions $f_\sigma({\bm p},t)$ are peaked in the narrow
region around a momentum ${\bm p}_0$. Then we obtain
\begin{eqnarray}\label{eq:f2}
\frac{d}{d t}[N_+(t)-N_-(t)]&=&-\Omega_{+}^{out}N_+(t)+
\Omega_{-}^{out}N_-(t)+2[\Omega_{+-}N_-(t)-\Omega_{-+}N_+(t)]\, ,\nonumber\\
&&\nonumber\\
\frac{d}{dt}[N_+(t)+N_-(t)]&=&-[\Omega_{+}^{out}N_+(t)+\Omega_{-}^{out}N_-(t)]\,
.
\end{eqnarray}
Here
\begin{eqnarray}\label{eq:notation1}
&&\Omega_+^{out}=\int\limits_{{\bm p}'\notin \Gamma} d{\bm
p}'[W_{++}({\bm p}',{\bm p}_0)+W_{-+}({\bm p}',{\bm p}_0)]\,
,\quad \Omega_-^{out}=\int\limits_{{\bm p}'\notin \Gamma} d{\bm
p}'[W_{--}({\bm p}',{\bm p}_0)+W_{+-}({\bm p}',{\bm p}_0)]\, ,
\nonumber\\
&&\Omega_{+-}=\int\limits_{{\bm p}'\in \Gamma} d{\bm
p}'W_{+-}({\bm p}',{\bm p}_0)\, , \quad
\Omega_{-+}=\int\limits_{{\bm p}'\in \Gamma} d{\bm p}'W_{-+}({\bm
p}',{\bm p}_0)\, ,\quad N_\pm(t)=\int\limits_{{\bm p}\in \Gamma}
d{\bm p}f_\pm({\bm p},t)\, .
\end{eqnarray}
The solution to Eq.(\ref{eq:f2}), with the initial condition
$N_+(0)=N_-(0)=N_0/2$, reads
\begin{eqnarray}\label{eq:solution}
&&N(t)=N_+(t)+N_-(t)=N_0\left[\cosh(\Omega
t)+\frac{\Omega_{+-}+\Omega_{-+}}{2\Omega}\sinh(\Omega
t)\right]\exp(-\Omega_{tot}t)\, ,
\nonumber\\
&&\nonumber\\
 &&P_B(t)=\frac{N_+(t)-N_-(t)}{N_+(t)+N_-(t)}=
\frac{\Omega_{+-}-\Omega_{-+}+\frac{1}{2}(\Omega_{-}^{out}-\Omega_{+}^{out})}
{\Omega+\frac{1}{2}(\Omega_{+-}+\Omega_{-+})\tanh(\Omega
t)}\tanh(\Omega t)\, ,
\end{eqnarray}
where $N(t)$ is the total number of particles in the beam,
$P_B(t)$ is the beam polarization, and
\begin{eqnarray}\label{eq:notation2}
&&\Omega_{tot}=\frac{1}{2}\left[\Omega_{+}^{out}+\Omega_{-}^{out}+
\Omega_{+-}+\Omega_{-+}\right]\, ,
\nonumber\\
&&\nonumber\\
 &&\Omega=\frac{1}{2}\biggl[(\Omega_{-}^{out}-\Omega_{+}^{out})^2+
 (\Omega_{+-}+\Omega_{-+})^2
 +2(\Omega_{-}^{out}-\Omega_{+}^{out})(\Omega_{+-}-\Omega_{-+})
 \biggr]^{1/2}
\, .
\end{eqnarray}
For the scattering angle $\vartheta >\theta_{acc}$, the momentum
transfer is much larger than $1/a_0$ ($a_0$ is the Bohr radius,
$\hbar=c=1$).  In this case scattering off hydrogen atom can be
considered as independent scattering off a free electron and a
free proton. The maximum scattering angle of a  proton on an
electron at rest is less than $\theta_{acc}$ for any storage ring.
Therefore, scattering on electrons at rest does not contribute to
$\Omega_\pm^{out}$. If the proton polarization of the hydrogen
target is absent (as in the scheme considered in
Ref.\cite{Rathmann05}), then $\Omega_{-}^{out}=\Omega_{+}^{out}$
and we have from Eq.(\ref{eq:solution})
\begin{eqnarray}\label{eq:example1}
P_B(t)=\frac{\Omega_{+-}-\Omega_{-+}}{\Omega_{+-}+\Omega_{-+}}
 \biggl[1-\exp[-(\Omega_{+-}+\Omega_{-+})\, t\,]\biggr]\,
\, .
\end{eqnarray}
If the proton polarization of the target is not small, then
$|\Omega_{-}^{out}-\Omega_{+}^{out}|$ is much larger than
$\Omega_{+-}$ and $\Omega_{-+}$. In this  case
\begin{eqnarray}\label{eq:example2}
P_B(t)=\tanh\biggl[\frac{t}{2}\,(\Omega_{-}^{out}-\Omega_{+}^{out})\biggr]\,
.
\end{eqnarray}

\section{Probabilities and cross sections for pp scattering}
Let us direct the  polar axis   along the unit vector ${\bm
\nu}={\bm p}_0/p_0$. Then the cross section of pp scattering
integrated over the azimuth angle and summed up over the final
polarization of both protons in the center-of-mass frame reads
\begin{eqnarray}\label{eq:sigma}
d\sigma=2\pi \sin\vartheta\,
d\vartheta\left\{F_0(\vartheta)+({\bm\zeta}_t\cdot{\bm\zeta}_b)F_1(\vartheta)+
({\bm\zeta}_t\cdot{\bm\nu})({\bm\zeta}_b\cdot{\bm\nu})[F_2(\vartheta)
-F_1(\vartheta)]\right\} \, ,
\end{eqnarray}
where ${\bm\zeta}_t$ and ${\bm\zeta}_b$ are the unit polarization
vectors of the protons from the target and the beam, respectively.
The functions $F_0(\vartheta)$, $F_1(\vartheta)$, and
$F_2(\vartheta)$ are
\begin{eqnarray}\label{eq:FFF}
F_0(\vartheta)&=&I_{0000}=\frac{1}{2}\left\{|M_1|^2
+|M_2|^2+|M_3|^2+|M_4|^2+4|M_5|^2\right\}\, ,\nonumber\\
F_1(\vartheta)&=&\frac{1}{2}I_{0000}\left\{A_{00nn}+\cos^2(\vartheta/2)\,A_{00mm}+
\sin^2(\vartheta/2)\,A_{00ll}+\sin(\vartheta)\,A_{00ml}\right\}\nonumber\\
&=&|M_5|^2+\mbox{Re}(M_1\,M_2^*)\, ,\nonumber\\
F_2(\vartheta)&=&I_{0000}\left\{\sin^2(\vartheta/2)\,A_{00mm}+
\cos^2(\vartheta/2)\,A_{00ll}-\sin(\vartheta)\,A_{00ml}\right\}\nonumber\\
&=&\frac{1}{2}\left\{|M_3|^2+|M_4|^2-|M_1|^2-|M_2|^2\right\}\, .
\end{eqnarray}
Here the observables $I_{0000}$, $A_{00mm}$, $A_{00nn}$,
$A_{00ll}$, $A_{00ml}$ and the helicity amplitudes $M_i$ are
defined as in Ref.\cite{Bystricky78}.

Let  ${\bm P}_T$ be the target polarization vector, and
${\bm\zeta}_T={\bm P}_T/ P_T$. We direct the quantization axis
along the unit vector ${\bm\zeta}_T$. Averaging the cross section
in Eq.(\ref{eq:sigma}) over  particles in the target, we obtain
for the quantities $\Omega_{\pm}^{out}$
\begin{eqnarray}\label{eq:Oout}
\Omega_{\pm}^{out}&=&nf\left\{\sigma_0^{out}\pm P_T\left[
\sigma_1^{out}+({\bm\zeta}_T\cdot{\bm
\nu})^2(\sigma_2^{out}-\sigma_1^{out})\right]\right\}\,
,\nonumber\\
\sigma_i^{out}&=&2\pi
\int\limits_{\theta_{acc}}^{\pi/2}d\vartheta\,\sin\vartheta\,F_i(\vartheta)\,
 .
\end{eqnarray}
Here $n$ is an  areal density  of the target and $f$ is a
revolution frequency. The function $P_B(t)$ in
Eq.(\ref{eq:solution}) contains $\Omega_{\pm}^{out}$ only in the
combination $\Omega_{+}^{out}-\Omega_{-}^{out}$. For
$|\sigma_2|>|\sigma_1|$, this difference is maximal  at
${\bm\zeta}_T
\parallel{\bm \nu}$. For $|\sigma_2|<|\sigma_1|$, the difference
is maximum at ${\bm\zeta}_T \perp{\bm \nu}$. In
Eqs.(\ref{eq:solution}) and (\ref{eq:notation2}), the quantities
$\Omega_{+-}$ and $\Omega_{-+}$, which are related to the
spin-flip transitions,   have the form
\begin{eqnarray}\label{eq:OPM}
\Omega_{-+}+\Omega_{+-}&=&nf\left\{\sigma_{2s}+({\bm\zeta}_T\cdot{\bm
\nu})^2\left( \sigma_{1s}-\sigma_{2s}\right)\right\}\,
,\nonumber\\
\Omega_{-+}-\Omega_{+-}&=&nf
P_T\left\{\sigma_{2d}+({\bm\zeta}_T\cdot{\bm \nu})^2\left(
\sigma_{1d}-\sigma_{2d}\right)\right\}\,
,\nonumber\\
\sigma_X&=&2\pi
\int\limits_{0}^{\theta_{acc}}d\vartheta\,\sin\vartheta\,G_X(\vartheta)\,
 .
\end{eqnarray}
In terms of the helicity amplitudes, the functions
$G_X(\vartheta)$ are
\begin{eqnarray}\label{eq:G}
G_{1s}(\vartheta)&=&\frac{1}{2}\biggl\{|2\cos(\vartheta/2)M_5+\sin(\vartheta/2)(M_1+M_3)|^2
+|2\sin(\vartheta/2)M_5+\cos(\vartheta/2)(M_4-M_2)|^2\nonumber\\
&&+\sin^2(\vartheta/2)|M_1-M_3|^2
+\cos^2(\vartheta/2)|M_2+M_4|^2\biggr\}\,
,\nonumber\\
G_{2s}(\vartheta)&=&\frac{1}{2}\biggl\{G_{1s}+
\cos^2(\vartheta/2)|M_1-M_3|^2+
\sin^2(\vartheta/2)|M_2+M_4|^2\biggr\}
\, ,\nonumber\\
G_{1d}(\vartheta)&=&\sin\vartheta\,\mbox{Re}\biggl[ M_5^*
(M_2+M_4+M_3-M_1)\biggr]+\sin^2(\vartheta/2)(|M_3|^2-|M_1|^2)\nonumber\\
&&+\cos^2(\vartheta/2)(|M_4|^2-|M_2|^2)\, ,\nonumber\\
G_{2d}(\vartheta)&=&\mbox{Re}\biggl\{\frac{1}{2}\sin\vartheta\,\biggl[
M_5^*
(M_2+M_4+M_3-M_1)\biggr]+M_2^*(M_1-M_3)\nonumber\\
&&+ \sin^2(\vartheta/2)(M_4^*M_1+M_2^*M_3)\biggr\}\, .
\end{eqnarray}
Each  amplitude $M_i$  can be represented as a sum
$M_i=M_i^{em}+M_i^{h}$ of a pure electromagnetic amplitude,
$M_i^{em}$, and the hadronic amplitude, $M_i^{h}$. The latter also
accounts for the electromagnetic interaction. The amplitudes
$M_i^{h}$ are not singular at small scattering angle $\vartheta$.
More precisely, at $\vartheta\to 0$, $M_{1,2,3}^{h}$ are nonzero
constant, $M_{4}^{h}\propto \vartheta^2$, and $M_{5}^{h}\propto
\vartheta$. In the nonrelativistic  limit, the amplitudes
$M_i^{em}$  pass into the amplitude $M_i^{C}$, which are the
matrix element of the operator $\hat M^{C}$ in the spin space
\cite{LL}
\begin{eqnarray}{\label{eq:MChat}}
\hat M^{C}&=&f(\vartheta)-\frac{1}{2}(1+
\bm{\sigma}_b\cdot\bm{\sigma}_t)f(\pi-\vartheta)\, ,\nonumber\\
f(\vartheta)&=&-\frac{\alpha
}{4vp\sin^2(\vartheta/2)}\exp\{-i(\alpha/v)\ln[\sin(\vartheta/2)]\}\,
,
\end{eqnarray}
where $v=p/m_p$ is the  proton velocity in the  center-of-mass
frame, $\bm\sigma$ are the Pauli matrices, $\alpha$ is the fine
structure constant. The presence of the spin operators in $\hat
M^C$ is completely due to the identity of protons. From
Eq.(\ref{eq:MChat}), we obtain for $M_i^{C}$
\begin{eqnarray}{\label{eq:MC}}
 M_1^{C}&=&\cos^2(\vartheta/2)f(\vartheta)+\sin^2(\vartheta/2)f(\pi-\vartheta)
 \, ,\nonumber\\
 M_2^{C}&=&-[\sin^2(\vartheta/2)f(\vartheta)+\cos^2(\vartheta/2)f(\pi-\vartheta)]\, ,
 \nonumber\\
M_3^{C}&=&\cos^2(\vartheta/2)[f(\vartheta)-f(\pi-\vartheta)]\,
,\nonumber\\
 M_4^{C}&=&\sin^2(\vartheta/2)[f(\vartheta)-f(\pi-\vartheta)]\, ,
\nonumber\\
  M_5^{C}&=&-\frac{1}{2}\sin\vartheta\,[f(\vartheta)-f(\pi-\vartheta)]\,
 .
 \end{eqnarray}

Using Eq.(\ref{eq:MC}), we can  consider  the interrelation
between the electromagnetic and hadronic contributions to
$\Omega_{\pm}^{out}$. If $\theta_{acc}\ll \alpha/(vpH)$,  $H$ is a
typical magnitude of the hadronic amplitudes, then the main
contribution to the cross section $\sigma_0$ in Eq.(\ref{eq:f})
comes from the integration region $\vartheta\sim\theta_{acc}\ll
1$, where $F_0(\vartheta)\simeq |f(\vartheta)|^2$. Thus, this
contribution has  pure electromagnetic origin:
\begin{equation}\label{eq:sigma0}
\sigma_0\approx \sigma_0^C=\pi\alpha^2/(vp\theta_{acc})^2\, .
\end{equation}
 For the electromagnetic part of the functions $F_{1,2}$, we have
$F_1^C=F_2^C=-\mbox{Re}[f^*(\vartheta)f(\pi-\vartheta)]$. The
corresponding contribution to $\sigma_{1,2}$ reads
\begin{equation}\label{eq:sigma12}
\sigma_1^C=\sigma_2^C=-\frac{\pi\alpha}{2vp^2} \sin\Psi\,,\quad
\Psi=\frac{\alpha}{v}\ln(2/\theta_{acc})\, .
\end{equation}
The interference terms $\sigma_i^{int}$ in $\sigma_{i}$ can be
estimated to the logarithmic accuracy as
\begin{eqnarray}\label{eq:sigmaint}
\sigma_0^{int}&=&-\frac{2\pi}{p}\left\{ \sin\Psi
\,\mbox{Re}[M_3^h(0)+M_1^h(0)]+(1-\cos\Psi)\,\mbox{Im}[M_3^h(0)+M_1^h(0)]\right\}\,
,\nonumber\\
\sigma_1^{int}&=&-\frac{2\pi}{p}\left\{ \sin\Psi
\,\mbox{Re}M_2^h(0)+(1-\cos\Psi)\mbox{Im}M_2^h(0)\right\}\,
,\nonumber\\
\sigma_2^{int}&=&-\frac{2\pi}{p}\left\{ \sin\Psi
\,\mbox{Re}[M_3^h(0)-M_1^h(0)]+(1-\cos\Psi)\,\mbox{Im}[M_3^h(0)-M_1^h(0)]\right\}\,.
\end{eqnarray}
We illustrate a scale of  different contributions, giving their
 numerical values corresponding to the parameters  of
the experiment \cite{Haeberli93}: $E_{lab}=23$MeV ,
$\theta_{acc}=8.8$mrad. Using the data base \cite{nn-online} for
the hadronic amplitudes $M_i^h$, we obtain
\begin{eqnarray}\label{eq:numbers}
\sigma_0&=&6444\,\mbox{mb}\, ,\quad
\sigma_0^{int}=-56\,\mbox{mb}\, ,\quad
\sigma_0^{C}=6357\,\mbox{mb}\,
,\nonumber\\
\sigma_1&=&-89\,\mbox{mb}\, ,\quad \sigma_1^{int}=39\,\mbox{mb}\,
,\quad \sigma_1^{C}=-1\,\mbox{mb}\,
,\nonumber\\
\sigma_2&=&-66\,\mbox{mb}\, ,\quad \sigma_2^{int}=66\,\mbox{mb}\,
,\quad \sigma_2^{C}=-1\,\mbox{mb}\,\,.
\end{eqnarray}
Recollect that $\sigma_i=\sigma_i^{em}+\sigma_i^{int}+\sigma_i^h$
and $\sigma_i^{em}\simeq\sigma_i^{C}$ in the nonrelativistic case.
Note that the numbers obtained from Eqs.(\ref{eq:sigma0}),
(\ref{eq:sigma12}) and (\ref{eq:sigmaint}) are in good agreement
with that in Eq.({\ref{eq:numbers}). The role of interference of
the hadronic and electromagnetic amplitudes is additionally
illustrated by Fig.\ref{figure}. In this figure, the function
$2\pi\sin\vartheta\,F_1(\vartheta)$ calculated with the use of the
full amplitudes $M_i=M_i^{em}+M_i^{h}$ (solid curve) is compared
to that obtained  using the hadronic amplitudes $M_i^{h}$ only
(dashed curve). Drastic modification of the function
$F_1(\vartheta)$ at small $\vartheta$, as compared to the hadronic
contribution, is due to interference (the pure electromagnetic
contribution is negligible).

\begin{figure}[ht]
\vspace{40pt} \centering
\includegraphics[height=165pt,keepaspectratio=true]{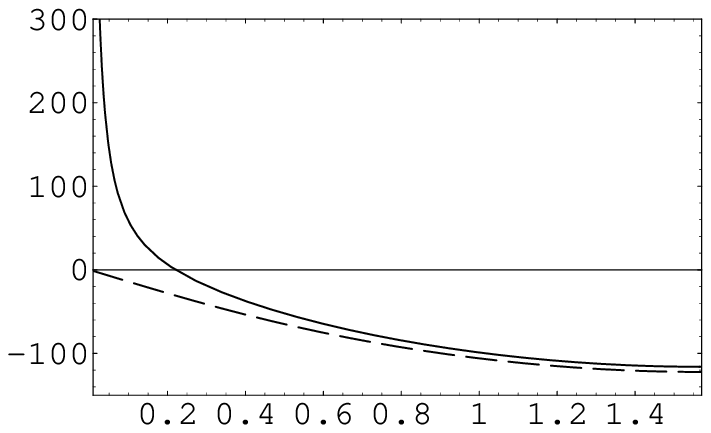}
\begin{picture}(0,0)(0,0)
 \put(-135,-10){\large $\vartheta$}
 \put(-280,35){\rotatebox{90}{\large$2\pi\sin\vartheta\,F_1(\vartheta)$(mb)}}
 \end{picture}
\caption{\label{figure}\it The function
$2\pi\sin\vartheta\,F_1(\vartheta)$ in units mb, calculated with
the use of full amplitudes (solid curve) and of hadronic
amplitudes only (dashed curve).}
\end{figure}

Let us consider the quantities $\sigma_{1s}$, $\sigma_{2s}$,
$\sigma_{1d}$, and $\sigma_{2d}$ (see Eq.(\ref{eq:OPM})), which
determine the functions $\Omega_{+-}$ and $\Omega_{-+}$. The
latter correspond to the spin-flip transitions at $\vartheta \leq
\theta_{acc}\ll 1$. Using the nonrelativistic electromagnetic
amplitudes from Eq.(\ref{eq:MC}), we obtain for the small-angle
asymptotics of the functions $G_X(\vartheta)$
\begin{eqnarray}\label{eq:sigpm}
G_{1s}&=&|M_2^h(0)|^2\, ,\quad
G_{2s}=\frac{1}{2}|M_2^h(0)|^2+\frac{1}{2}|M_1^h(0)-M_3^h(0)|^2\,,
\nonumber\\
G_{1d}&=&-|M_2^h(0)|^2\, ,\quad
G_{2d}=\mbox{Re}\{M_2^{h*}(0)[M_1^h(0)-M_3^h(0)]\}\,.
\end{eqnarray}
This  contribution is purely hadronic, and the corresponding cross
sections can be estimated as $\sigma_{s}\sim\sigma_{d}\sim
\theta_{acc}^2\sigma_{1}^h$, being negligibly small.

If we take into account the first relativistic correction to the
electromagnetic amplitudes (see, e.g., \cite{BLP}), then  the
additional term appears in the functions $G_{1s}$ and $G_{2s}$:
\begin{equation}
\delta G_{1s}=2\delta G_{2s}\simeq \frac{1}{8}\left[\frac{\alpha
(4\kappa+3)}{m_p\vartheta}\right]^2\, ,
\end{equation}
where $\kappa=1.79$ is the  anomalous magnetic moment of the
proton. Though the corresponding cross sections is logarithmically
enhanced, it is negligibly small being proportional to
$\alpha^2/m_p^2$. Thus, scattering events, where projectiles stay
in the beam, do not lead to the  beam polarization.

\section{Pure electron target}
The time dependence of the beam polarization is described again by
Eq.(\ref{eq:solution}) , but with another expressions for the
cross sections in Eqs.(\ref{eq:Oout}) and (\ref{eq:OPM}), which
correspond now to the pure electromagnetic electron-proton
interaction. The cross section accounting for the polarizations of
the initial and final proton and electron is well known (see,
e.g.,\cite{Scofield1966}).

The functions $\Omega_+^{out}$ and $\Omega_-^{out}$ are  related
to  the scattering events, where  particles go out from the beam.
Below we  consider  head-on collisions of electrons and protons in
the laboratory frame. For any storage ring, $\theta_{acc}\gg
m_e/m_p\approx 0.5\mbox{mrad}$. Then the particle loss occurs at
$p_e>p^{out}$, where
\begin{equation}
p^{out}=\frac{p_p\theta_{acc}m_p}{\varepsilon_p+p_p}\, ,
\end{equation}
$\varepsilon_p=\sqrt{p_p^2+m_p^2}$ is the energy and $p_p$ is the
momentum of a proton,   $ p_e$ is the momentum of an electron. For
$p_e< p_p\,m_p/(\varepsilon_p+p_p+m_p)$, there is the maximal
scattering angle $\theta_{max}$,
\begin{equation}
\theta_{max}=\arcsin\left[\frac{\varepsilon_pp_e+\varepsilon_ep_p}
{m_p(p_p-p_e)}\right]\, ,
\end{equation}
where $\varepsilon_e$ is the electron energy. In particular,
$\theta_{max}=m_e/m_p$ for $p_e=0$. For $p_e>p^{out}$, the cross
sections $\sigma_0^{out}$, $\sigma_1^{out}$, and $\sigma_2^{out}$
in Eq.(\ref{eq:Oout}) can be estimated as
\begin{equation}
\sigma_0^{out}\sim \frac{4\pi\alpha^2}{(p_p\theta_{acc})^2}
\left(\frac{\varepsilon_e\varepsilon_p+p_ep_p}
{\varepsilon_ep_p+\varepsilon_pp_e}\right)^2 \, ,\quad
\frac{\sigma_1^{out}}{\sigma_0^{out}}\sim
\frac{m_em_p(p_p\theta_{acc})^2}{(\varepsilon_e\varepsilon_p+p_ep_p)^2}\,,\quad
\frac{\sigma_2^{out}}{\sigma_0^{out}}\sim
\frac{(p_p\theta_{acc})^2}{\varepsilon_e\varepsilon_p+p_ep_p}\,.
\end{equation}
For the time $t\sim\Omega_{tot}^{-1}$, when $N(t)/N(0)$ is not too
small (see Eq.({\ref{eq:solution})), the ratios
$\sigma_1^{out}/\sigma_0^{out}$ and
$\sigma_2^{out}/\sigma_0^{out}$ give the estimations of the beam
polarization $P_B(t)$ for ${\bm\zeta}_T \perp {\bm \nu}$ and
${\bm\zeta}_T
\parallel {\bm \nu}$, respectively. These ratios  are
maximal for $p_e\gtrsim p^{out}$, so that $P_B<m_e/m_p$ for
${\bm\zeta}_T \perp {\bm \nu}$, and $P_B<p_p\theta_{acc}/m_p$ for
${\bm\zeta}_T\parallel {\bm \nu}$. In both cases $P_B$ is too
small, and  the mechanism using the loss of particles due to
proton scattering on  polarized electrons (filtering) does not
work for any parameters of the electron beam.

Let us consider now the mechanism of the  polarization buildup
without loss of the particles, which is due to the spin-flip
transitions. Starting with the general expressions for the cross
section of polarized electron-proton scattering,
\cite{Scofield1966}, we find that the cross  sections $\sigma_s$
and $\sigma_d$ in Eq.(\ref{eq:OPM}) are maximal at small relative
velocity of electron and proton. In this case, we estimate
$\sigma_{d}\sim 16\pi\mu_p^2\sim 10^{-3}$mb , $\mu_p$ is the
proton magnetic moment. As compared with $\sigma_{d}$, the cross
section $\sigma_{s}$ is enhanced by some logarithmic factor, which
can not change essentially the conclusion on its smallness. We
emphasize that the cross sections are small because the spin-flip
amplitude does not contain a  term $\propto 1/\vartheta^2$ at
$\vartheta\to 0$, contrary to the amplitude without the spin-flip
transition where such term (Coulomb term) is present. Thus, for a
pure electron target, the mechanism of  polarization buildup based
on the spin-flip transitions also does not work.

In conclusion, it is demonstrated that, for $\theta<\theta_{acc}$
(a proton remains in the beam), the polarization buildup is
completely due to the spin-flip transitions. The corresponding
cross sections turn out to be  negligibly small for both
proton-proton and proton-electron scattering. For a  pure electron
target, filtering mechanism  also does not provide  a noticeable
polarization. Evidently, these statements are valid for the
antiproton beam as well. Thus, the filtering method using a
hydrogen gas target with the proton polarization seems to be the
most promising way to polarize stored antiprotons.

\section*{ACKNOWLEDGMENTS}
We are grateful  to  I.A.Koop, V.V.Parkhomchuk, Yu.M.Shatunov,
A.N.Skrinsky, and D.K.Toporkov for stimulating discussions.
 This work was supported in part by  RFBR Grants  03-02-16510
 and 05-02-16079.

\end{document}